\documentclass[reprint, amsmath,amssymb, aps, superscriptaddress, floatfix]{revtex4-1}

\usepackage[utf8]{inputenc}
\usepackage{graphicx}
\usepackage{dcolumn}
\usepackage{bm}
\usepackage{color}
\usepackage{units}
\usepackage{dsfont}
\usepackage{float}
\usepackage{hyperref}
\usepackage{xcolor}
\usepackage{amssymb}

\newcommand{\bra}[1]{\left\langle #1 \right|}
\newcommand{\ket}[1]{\left| #1 \right\rangle}
\newcommand{\abs}[1]{\left| #1 \right|}
\newcommand{\braket}[2]{\left\langle {#1{\left| \vphantom{#1 #2} \right.} #2} \right\rangle}

\begin{document}

\title{Underwater quantum communication over a 30-meter flume tank}

\author{Felix Hufnagel}
\affiliation{Department of physics, University of Ottawa, Advanced Research Complex, 25 Templeton Street, K1N 6N5, Ottawa, ON, Canada}

\author{Alicia Sit}
\affiliation{Department of physics, University of Ottawa, Advanced Research Complex, 25 Templeton Street, K1N 6N5, Ottawa, ON, Canada}

\author{Fr\'ed\'eric Bouchard}
\email{frederic.bouchard@nrc-cnrc.gc.ca}
\affiliation{Department of physics, University of Ottawa, Advanced Research Complex, 25 Templeton Street, K1N 6N5, Ottawa, ON, Canada}
\affiliation{National Research Council of Canada, 100 Sussex Drive, Ottawa, Ontario K1A 0R6, Canada}

\author{Yingwen Zhang}
\affiliation{National Research Council of Canada, 100 Sussex Drive, Ottawa, Ontario K1A 0R6, Canada}

\author{Duncan England}
\affiliation{National Research Council of Canada, 100 Sussex Drive, Ottawa, Ontario K1A 0R6, Canada}

\author{Khabat Heshami}
\affiliation{Department of physics, University of Ottawa, Advanced Research Complex, 25 Templeton Street, K1N 6N5, Ottawa, ON, Canada}
\affiliation{National Research Council of Canada, 100 Sussex Drive, Ottawa, Ontario K1A 0R6, Canada}

\author{Benjamin J. Sussman}
\affiliation{Department of physics, University of Ottawa, Advanced Research Complex, 25 Templeton Street, K1N 6N5, Ottawa, ON, Canada}
\affiliation{National Research Council of Canada, 100 Sussex Drive, Ottawa, Ontario K1A 0R6, Canada}

\author{Ebrahim Karimi}
\affiliation{Department of physics, University of Ottawa, Advanced Research Complex, 25 Templeton Street, K1N 6N5, Ottawa, ON, Canada}
\affiliation{National Research Council of Canada, 100 Sussex Drive, Ottawa, Ontario K1A 0R6, Canada}

\begin{abstract}
Underwater quantum communication has recently been explored using polarization and orbital angular momentum. Here, we show that spatially structured modes, e.g., a coherent superposition of beams carrying both polarization and orbital angular momentum, can also be used for underwater quantum cryptography. We also use the polarization degree of freedom for quantum communication in an underwater channel having various lengths, up to $30$ meters. The underwater channel proves to be a difficult environment for establishing quantum communication as underwater optical turbulence results in significant beam wandering and distortions. However, the errors associated to the turbulence do not result in error rates above the threshold for establishing a positive key in a quantum communication link with both the polarization and spatially structured photons. The impact of the underwater channel on the spatially structured modes is also investigated at different distances using polarization tomography.
\end{abstract}

\maketitle


\section{Introduction}
The development of quantum computers and their ability to factor large prime numbers through Shor's algorithm, poses a threat to modern communication security~\cite{shore94}. Since the realization of the first protocol by  Bennett and Brassard in 1984 (BB84), quantum key distribution (QKD) has become the most actively researched solution for secure communication in a post-quantum world~\cite{bennett1984quantum}.
At present, quantum communication is a research field reaching maturity with commercial devices available for optical fibre QKD, experimental demonstrations of satellite-to-ground channels~\cite{vallone:15,liao:17}, and new protocols being developed to improve key rates and security~\cite{pirandola2019}. There are two primary quantum communication instances currently investigated: quantum communication channels and quantum communication protocols~\cite{Leuchs:20,ekert1991,bennett:92}. Channels must be experimentally investigated to demonstrate the transmission fidelity of quantum states, while new protocols bring advantages in terms of security, error tolerance, and key rates. 

Many different protocols have been developed, some of which take advantage of high-dimensional quantum states~\cite{Bouchard2018experimental}. Although quantum information is typically encoded using one of the different photonic degrees of freedom (such as polarization, time-bin, position and transverse momentum), multiple degrees of freedom can be used, forming structured states of light. The coherent combination of polarization and orbital angular momentum (OAM) is one example of structured light, with numerous applications in microscopy, optical tweezers, classical communication, and quantum information~\cite{roadmap:16}. 
QKD protocols performed across free-space channels have been implemented, for example, with polarization~\cite{buttler:98}, time-bin~\cite{jin:19},  OAM~\cite{mirhosseini:15}, and structured photons~\cite{vallone2014free,Sit:17}. Likewise, there have been similar demonstrations in fibre channels~\cite{muller:93,marand:95,sit:18,xavier:20}. Though two-dimensional qubit protocols are the most commonly implemented, there are advantages to high-dimensional communication in both noise tolerance~\cite{ecker2019overcoming} and bit-rate~\cite{islam2017provably} which motivates the study of structured states~\cite{cozzolino2019}. 
\begin{figure*}[t]
\centering
\includegraphics[width=0.9\textwidth]{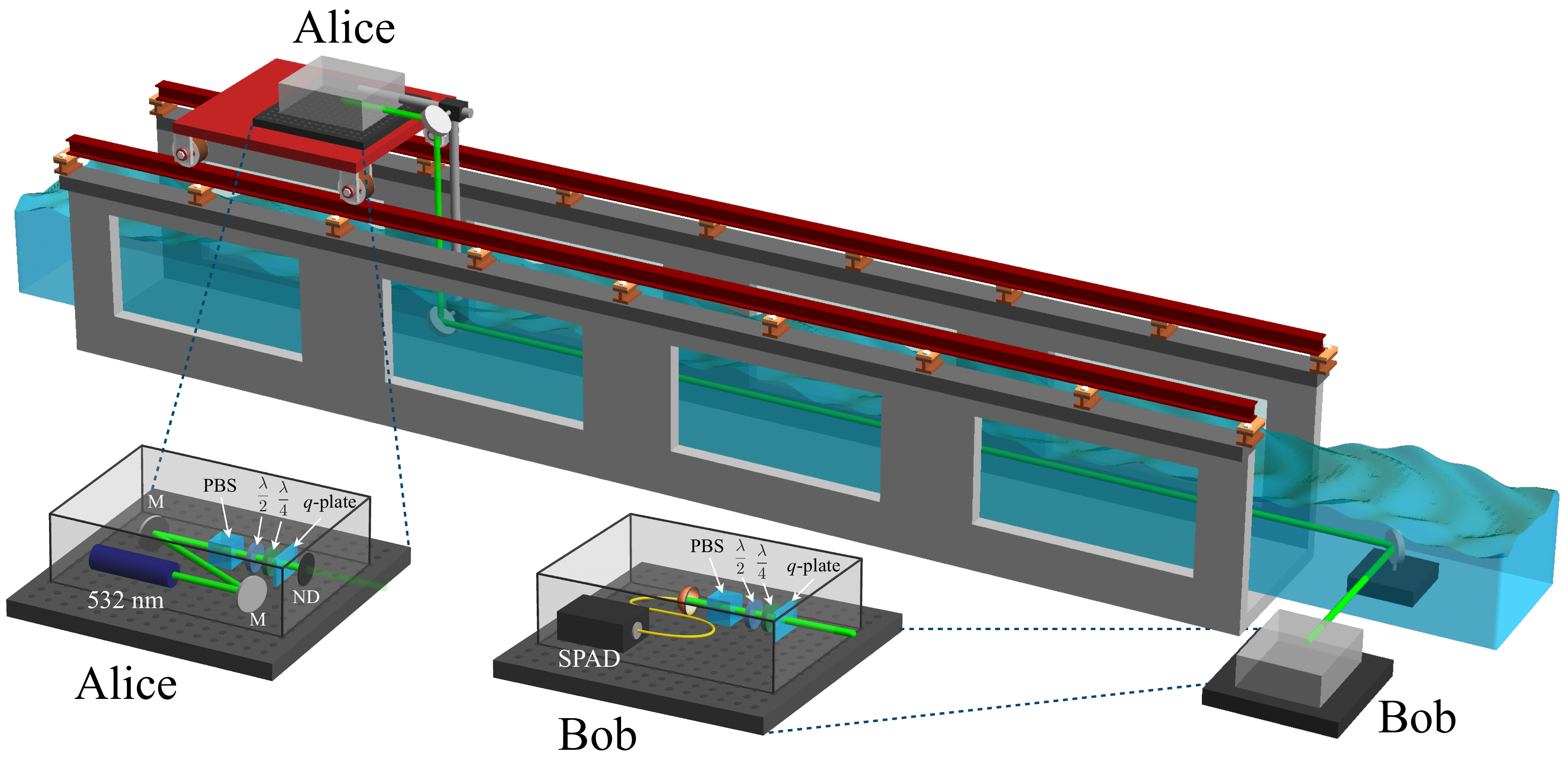} 
\caption{\textbf{Experimental setup.} The underwater channel was created using a flume 1~m deep and wide, and 50~m long. The sender setup, Alice, was placed on a trolley which was able to run the length of the flume. Alice's setup consisted of a 532~nm diode laser as the source, and a $\lambda/2$ and $\lambda/4$ waveplate, as well as a $q=1/2$-plate for the case of the vector vortex modes experiment. The beam was sent from the sender, through a glass tube which entered the water thus avoiding the wavy air-water interface. The beam was directed, from a mirror placed in the tank, through a window on the side of the flume to Bob's setup. Bob's setup, which mirrors Alice's setup, consisted of a $q=1/2$-plate followed by a $\lambda/2$ and $\lambda/4$ waveplate and a polarizing beam splitter (PBS). Photons are then coupled to a single-mode optical fibre and detected by a single photon avalanche diode (SPAD). In order to perform polarization tomography, the reduced Stokes parameters were measured by means of a $\lambda/2$ and $\lambda/4$ waveplate, polarizing beam splitter (PBS) followed by a CCD camera.}
\label{fig:fig1}
\end{figure*}

Recently, there have been several studies demonstrating quantum communication tasks in underwater environments~\cite{bouchardPool:18,zhao:19,hufnagel:19,hu:19}, motivated by the goal of secure communication between submersibles, research vessels, and surface vehicles. The absorption of water at radio frequencies has resulted in acoustic technology being the dominant form of communication for the last 100 years. There exists a transparency window between 400~nm and 500~nm, giving the possibility of optical and quantum communication at these wavelengths. Theoretical works predict a maximum practical quantum channel length of 300~m at 418~nm in clear conditions~\cite{sullivan:63}. Though this is a short distance as compared to the current fibre and free-space links, it is enough for communication between submersibles and surface vehicles. Thus far, the feasibility tests of QKD with polarization encoding has been performed through a 3-m tube of water~\cite{zhao:19}, and through an indoor testing channel of 55~m~\cite{hu:19}. However much like in air, uncontrolled underwater channels are prone to the effects of turbulence---caused by local changes in the refractive index of water from temperature variations---which displaces and distorts a transmitted beam of light. Recent studies have explored the effects of underwater turbulence on QKD protocols and spatial modes of light in uncontrolled~\cite{bouchardPool:18} and natural~\cite{hufnagel:19} water channels. However, the security analysis of different protocols is not considered in these previous studies. The high loss nature of underwater channels makes the optimization of the protocol necessary to achieve maximum distances and key rates. 
Here, we investigate an underwater channel at various lengths up to 30~m. The turbulence impact on the channel is observed through polarization tomography of the spatial modes, showing mode degradation upon propagation through the water. The BB84 protocol is tested using both polarization states and two-dimensional spatially structured modes. Key rate analysis is performed showing how optimization of the protocol's parameters is important in the high loss scenarios observed in underwater channels. Moreover the variable channel length gives us a real-world look into the scaling of such quantum protocols, and the impact on the key rates. 

\section{Experiment}
The experimental setup consists of a sender (Alice) and receiver (Bob), as well as the flume tank which provides our underwater channel, see Fig.~\ref{fig:fig1}. The flume is 1-meter wide and 1-meter tall; its full length was 50~m. On the top of the flume was a track with a mounted trolley which can travel the length of the channel. Alice's setup was placed on the trolley such that the channel length could be adjusted from 1~m to the full 50~m. A periscope was hung from Alice's setup so that, with proper beam alignment, the trolley could be moved up and down the length of the flume while maintaining beam coupling to a single mode optical fibre at the receiver. The sender consisted of two configurations for changing between sending polarization states and sending the spatially structured modes, e.g., vector vortex modes~\cite{cardano:12}. For sending polarization states, a 532~nm diode was sent to a polarizing beamsplitter (PBS), followed by a half-wave plate ($\lambda/2$) and a quarter-wave plate ($\lambda/4$) to allow any polarization state to be generated. Before being transmitted across the channel, the appropriate neutral density (ND) filters attenuated the beam to achieve a mean photon number of 0.1 photons per nanosecond. The vector vortex modes were generated by adding a $q$-plate with a topological charge of $q=1/2$ after the polarization optics. At the receiver, the beam is initially collected by a 3-inch lens so that the whole beam is gathered even with slight beam wandering due to the underwater turbulence. The beam then passes through a $\lambda/2$ and $\lambda/4$ waveplate and then a PBS to project on a particular polarization state. The beam is then coupled to a single mode fibre connected to a single photon avalanche diode (SPAD) detector. To detect the vector vortex states, a $q$-plate with charge $q=1/2$ is placed at the receiver before the polarization optics. In this way, the detection system mirrors the generation system. The setup is initially optimized with the trolley placed at 1~m from the receiver. 

Linear polarization states are used to establish the polarization quantum communication channel. The two mutually unbiased bases (MUB) for encryption are $\ket{\psi_i} \in \{\ket{H},\ket{V}\}$ and $\ket{\phi_j}\in \{\ket{A}=\tfrac{1}{\sqrt{2}}(\ket{H}+\ket{V}),\;\;\ket{D}=\tfrac{1}{\sqrt{2}}(\ket{H}-\ket{V})\}$, respectively. MUB possess the property that a projection made on the incorrect basis results in no information gained about the state of the photon, i.e. a probability of $1/2$ for each state $\abs{\braket{\psi_i}{\phi_j}}^2=1/2$. A full probability-of-detection matrix is determined by sending each of the four polarization states and subsequently performing projective measurements of the four states at the receiver setup. This probability-of-detection matrix gives us the quantum bit error rate (QBER) which is used to calculate the key rate that can be achieved with the underwater channel. The uniformly polarized photons, i.e. $\ket{\psi_i}$ and $\ket{\phi_j}$, should remain largely unaffected by the turbulence because water is not a birefringent medium. Thus, the introduced errors will be negligible for short distances. The length of the channel and level of turbulence will, however, introduce more losses as the photons are not able to be gathered at the receiver. This will have some effect on the measured error rate since the losses in the signal will give more weight to the dark counts and background noise as the distance increases. 
\begin{table}[htb]
	\centering
	\caption{\textbf{Quantum bit error rate and key rates for polarization BB84.} QBER and key rates per sifted photon for the polarization BB84 protocol are measured at different propagation distances. Note that in principle, the polarisation states can be modulated and detected at 100 MHz, and thus the actual key rate is \textit{key rate per sifted photons}~$\times$~\textit{modulation rate}~$\times$~\textit{sifting efficiency}.\newline}
	\begin{tabular}{@{} |p{2.8cm}||p{1.2cm}|p{1.2cm}|p{1.2cm}|p{1.2cm}| @{}} \hline
		\textbf{distance} & 0.5~m & 10.5~m & 20.5~m & 30.5~m  \\ \hline \hline
		\textbf{QBER (\%)} & 0.27 & 0.74  & 3.7 & 0.96 \\ \hline
		\textbf{key rate} & 0.94 & 0.87 & 0.54 & 0.84 \\ \hline
	\end{tabular}
	\label{table1}
\end{table}\\
The QBER and key rates for the different distances are given in Table~\ref{table1}. The results show the QBER increasing slightly as the channel length is increased. The primary source of these increased errors is the losses that result from the turbulence at the longer distances. The type of experiment we performed involves projecting on one of the polarization states and recording the number of counts during a set time period. The unpredictable nature of the turbulence in the channel results in some periods of time having larger losses due to beam wandering and spatial distortions. The beam wandering and spatial distortions, of course, increase with distance, but are effectively random and difficult to predict in our uncontrolled environment. This effect, which is random, results in the high error rate observed for the 20.5~m channel in comparison with the other channel lengths.\\

The structured photon states, here, vector vortex modes, are generated at the sender, and detected at the receiver by adding $q$-plates to the setup---details are given in the caption of Fig.~\ref{fig:fig1}. $q$-plates are patterned liquid crystal devices which introduce a polarization dependent geometric phase across the plate when it is illuminated with circularly polarized beams~\cite{marrucci2006optical,larocque2016}. The transformation of a perfectly tuned $q=1/2$-plate in the circular polarization basis, left-circular $\mathbf{e}_L$ and right-circular $\mathbf{e}_R$ states, can be described as,
\begin{equation}
\begin{bmatrix}
\mathbf{e}_L \\
\mathbf{e}_R \\
\end{bmatrix}\;
\underrightarrow{\text{QP}}\;
\begin{bmatrix}
\mathbf{e}_R\, e^{i\phi } \\
\mathbf{e}_L\, e^{-i\phi} \\
\end{bmatrix}.
\end{equation}
As we can see from the above equation, the $q=1/2$-plate will act on circular polarization states by converting left circularly polarized photons to right circularly polarized photons with OAM of $\ell = +1$. Similarly, incident photons with right circular polarization will be converted to left circular with OAM of the opposite sign, i.e. $\ell=-1$. The states used for our structured QKD protocol are created by sending linear polarization states, $\left\{\left\{\ket{H},\ket{V}\right\},\left\{\ket{A},\ket{D}\right\}\right\}$, onto the $q=1/2$-plate. This results in a state that has a spatially dependent polarization in the form of radial, azimuthal, clockwise and counterclockwise sink topology, all possessing polarization topological charge of $+1$~\cite{cardano:12}. This is opposed to the case when a circular polarization state is sent to the $q$-plate resulting in uniform conversion to the opposite circular polarization and the addition of an OAM of $\ell=\pm 1$. The protocol we implemented uses the structured modes as a 2-dimensional Hilbert space. The first MUB contained the radial and azimuthal states, i.e.,
\begin{equation}\label{vectormode1}
\ket{\Psi_i} \in\left\{ \frac{(\ket{L,-1}+\ket{R,+1})}{\sqrt{2}},\frac{(\ket{L,-1}-\ket{R,+1})}{\sqrt{2}}\right\},
\end{equation}
where $\ket{\pi,\ell}$ indicates the photons with polarisation and OAM states of $\ket{\pi}$ and $\ket{\ell}$, respectively. The states for the second MUB were the polarisation patterns in the form of clockwise and counterclockwise sinks (vortex) which were generated by sending the $\ket{A}$ and $\ket{D}$ polarization states onto the $q=1/2$-plate, i.e.,
\begin{equation}\label{vectormode2}
\ket{\Phi_i} \in \left\{\frac{(\ket{L,-1}+i\ket{R,+1})}{\sqrt{2}},\frac{(\ket{L,-1}-i\ket{R,+1})}{\sqrt{2}}\right\}.
\end{equation}
\begin{table}[htb]
	\centering
	\caption{\textbf{Quantum bit error rate and key rates for a two-dimensional BB84 using vector vortex beam.} QBER and key rates per sifted photon for the vector vortex modes protocol are measured at different propagation distances. Note that in principle the mode generation and detection can be achieved by polarisation modulators at 100 MHz before (after) q-plates, and thus the actual key rate is \textit{key rate per sifted photons}~$\times$~\textit{modulation rate}~$\times$~\textit{sifting efficiency}.\newline}
	\begin{tabular}{@{} |p{2.8cm}||p{1.2cm}|p{1.2cm}|p{1.2cm}| @{}} \hline
		\textbf{distance} & 1.5~m & 5.5~m & 10.5~m \\ \hline \hline
		\textbf{QBER (\%)} & 1.44 & 3.4  & 1.0 \\ \hline
		\textbf{key rate } & 0.79 & 0.57 & 0.84 \\ \hline
	\end{tabular}
	\label{table2}
\end{table}
\begin{figure*}[t]
    \centering
    \includegraphics[width=1.8\columnwidth]{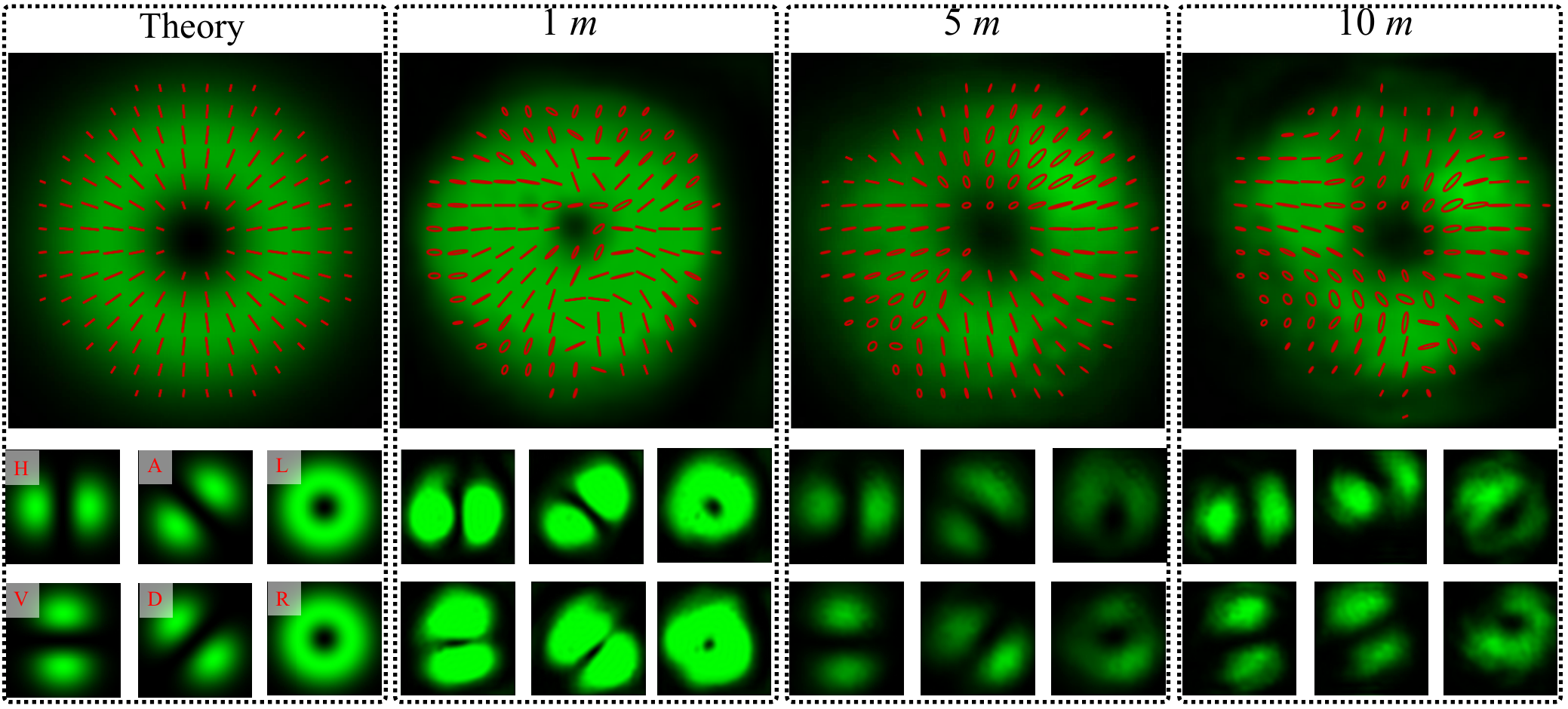}
    \caption{\textbf{Polarization tomography of a radially polarized beam for different channel lengths.} The reconstructed spatial polarization profile of the radial polarization mode is shown (top) for the theoretical as well as experimentally reconstructed beams propagating through the channel length of 1~m, 5~m, and 10~m. The intensity profile is shown in green, while the orientation of the polarization is shown in the red ellipses. The reconstructed spatial polarization profile was found using polarization tomography. The reduced stokes parameters were calculated for each pixel using the polarization measurements, \{H,V,D,A,L,R\}, shown in the lower insets.}
    \label{fig:fig2}
\end{figure*}

The combination of polarization, e.g., $\left\{\ket{H},\ket{V}\right\}$ and OAM, e.g., $\{-1,+1\}$ can be used as a 4-dimensional Hilbert space for high-dimensional~QKD~\cite{Sit:17}. To avoid errors assigned to high-dimensional generation and detection schemes~\cite{Bouchard2018experimental,zhu2019high}, we compare the 2-dimensional spatial modes with the polarization states. The vector vortex modes in Eqs. (\ref{vectormode1}) and (\ref{vectormode2}) were used to establish a quantum channel at 1, 5, and 10 meter distances. The difficulty with alignment of these spatial modes restricted us from coupling to single mode fiber for distances longer than 10 meters. The QBER and consequent key rates for these channels are given in Table~\ref{table2}. 

We also study the effect of underwater turbulence on the spatial profile of vector vortex modes propagating through different channel lengths. We perform polarization tomography by measuring the intensity of the structured beam, using a CCD camera, after passing a polarizer for \{H,V,A,D,R,L\}. From these measurements, the polarization is reconstructed pixel-by-pixel. The state sent across the channel is the radial polarization state.  This is created by sending a vertical linearly polarized Gaussian beam onto the $q=1/2$ plate. The resultant spin-orbit state is $(\ket{L,-1}+\ket{R,+1})/\sqrt{2}$. The theoretical polarization profile as well as the measured profiles for 1, 5, and 10 meters is shown in Fig.~\ref{fig:fig2}. The spatial profile can be seen to degrade as the channel length increases. This is associated to turbulence in the water channel that increases by the length~\cite{bouchardPool:18,hufnagel:19}. The phase distortion associated to the turbulence includes not only tip-tilt aberrations, but also higher-order effects such as astigmatism to the spatial profile of the beam. The aberrations in the underwater scenario are slower than those observed in air free-space, and thus the astigmatism aberrations are more easily seen as opposed to the scintillation often observed in high turbulence cases in the air. 

\section{Key Rate Analysis}
The initial BB84 proposal is an ideal protocol in which real single photons guarantee that there are not multiple photons in any given state which could open a door for an evesdropping attack, i.e., photon number splitting attack~\cite{brassard2000limitations}. For the BB84 protocol the number of bits that can be gained from each signal is,
\begin{equation}
G = \frac{1}{2}Q(1-2H(e)),
\end{equation}
where $Q$ is the gain and $H(e)=-e\log_2{(e)}-(1-e)\log_2{(1-e)}$ is the Shannon entropy for the error rate $e$. This ideal BB84 key rate requires some assumptions that are often not valid in practical QKD situations. The first of these assumption is the use of real single photons, i.e., heralded pairs, as opposed to an attenuated laser source. An attenuated laser source is a much more practical implementation, as photon pairs generated through spontaneous parametric down conversion (SPDC) are not deterministic. When using an attenuated source, the main point of weakness to consider is the possibility of more than one photon in a pulse. Other practical factors to consider are the effect of imperfect detectors and channel losses. These are brought into consideration in modern security proofs done for practical QKD protocols. We implement the decoy state protocol in our key rate analysis~\cite{lo2005decoy}. In any protocol using an attenuated laser source, one must choose the pulse mean photon number ($\mu$). This $\mu$ is taken below 1 such that the Poisson distribution gives a low probability of a 2-photon state. The density matrix of this signal state is given as, 
\begin{table*}[htb]
	\centering
	\caption{\textbf{Experimental parameters of the underwater polarization channel.} \newline}
    \begin{tabular}{c|c|c|c|c|c} \hline\hline
		\textbf{Parameter} & Dark Counts & Source Rep Rate & Detector Efficiency & Bob's Detection Efficiency & Channel Loss ($\alpha$) \\ \hline
		\textbf{Flume Result} & 300~Hz & $10^9$~Hz  & 0.6 & 0.188 & 0.57 dB/m \\ \hline \hline
	\end{tabular}
	\label{table3}
\end{table*}
\begin{equation}
    \rho = \sum_{i=0}^{\infty}\frac{\mu^i}{i!}e^{-\mu}\ket{i}\bra{i}.
\end{equation}
Here $\ket{i}$ is the Fock state, denoting $i$ number of photons. In the decoy state protocol, we include another state, the decoy state, which has a different mean photon number ($\nu$). The key rate for the decoy state protocol is given by,
\begin{equation}
    K=\frac{1}{2}\{-Q_{\mu}f(E_{\mu})H(E_\mu)+Q_1(1-H(e_1))\},
\end{equation}
where $Q_{\mu}$ is the gain of the total signal state sent by Alice, $E_\mu$ is the signal state QBER, $Q_1$ is the gain of the single photon state, $e_1$ is the error rate of the single photon state, and $f$ is the error correction efficiency. $Q_1$ and $e_1$ are respectively lower and upper bounded by,
\begin{eqnarray}
Q_1 &\geq& \frac{\mu^2 e^{-\mu}}{\mu \nu - \nu^2} \left(Q_\nu e^\nu - Q_\mu e^\mu \frac{\nu^2}{\mu^2} - \frac{\mu^2 - \nu^2}{\mu^2} Y_0 \right),
\end{eqnarray}
\begin{eqnarray}
e_1 \leq \frac{E_\nu Q_\nu e^\nu - Y_0/2}{Q_1 \nu/\left( \mu e^{-\mu} \right)}.
\end{eqnarray}
\begin{figure}[!htb]
	\includegraphics[width=1\columnwidth]{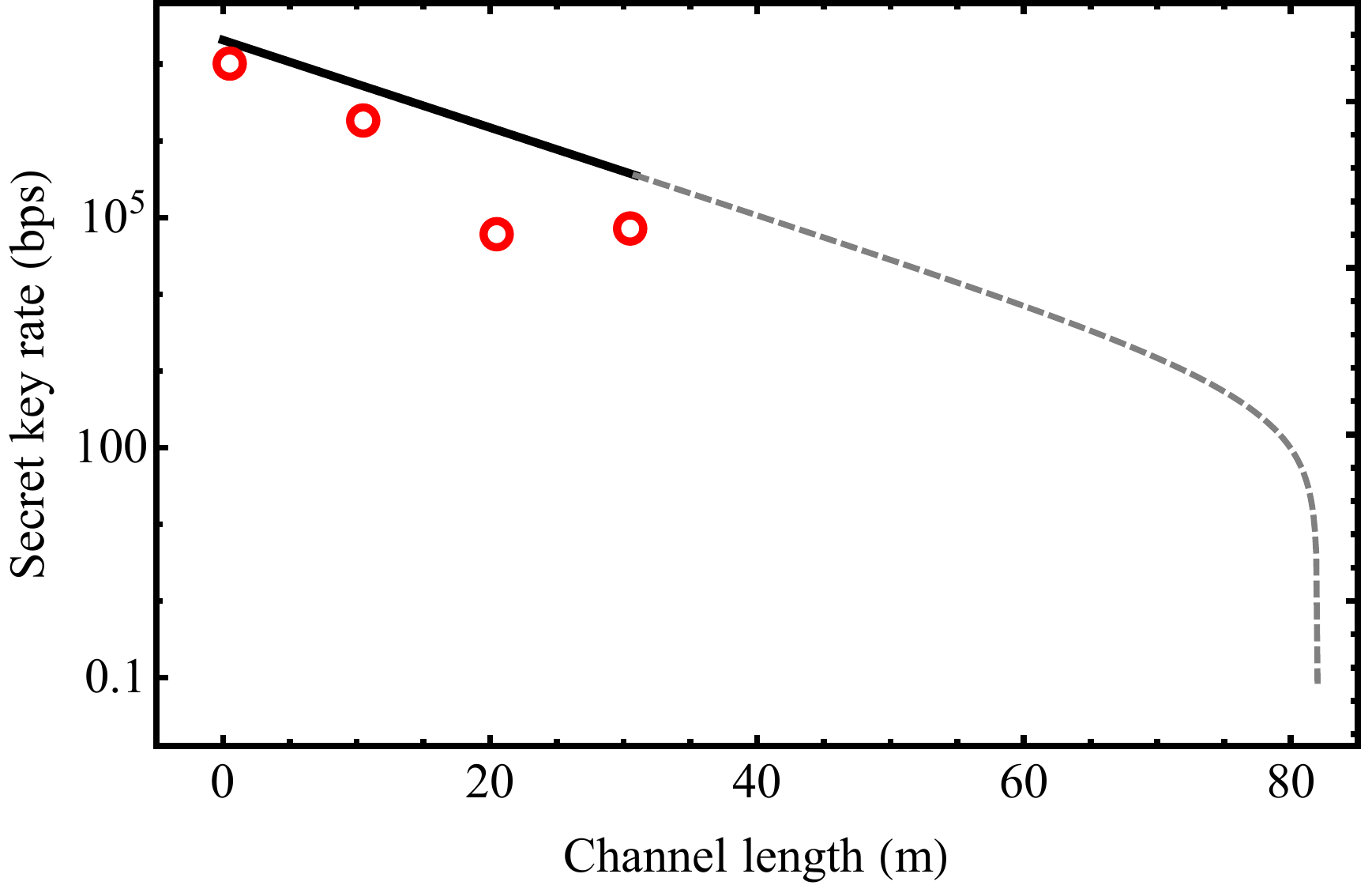}
	\caption{\textbf{Secret key rate for the underwater polarization channel.} The optimal secret key rate for the underwater channel with the measured attenuation is shown by the black curve. The experimental points (red) show the key rates achieved with the quantum bit error rates measured for the 0.5~m, 10.5~m, 20.5~m, 30.5~m channels. The experimental points, though necessarily lower than the zero-QBER theory, does follow the theoretical line closely, giving support to the prediction that 80 meters can be achieved. The dashed line shows the expected optimal key rate for channel lengths longer than 30~m. The optimization calculation is performed using the values stated in Table~\ref{table3}.}
	\label{fig:fig4}
\end{figure}

Here, $Y_0$ is the background rate per pulse, $Q_\nu$ is decoy state gain and $E_\nu$ is the decoy state QBER. We can optimize the values of $\mu$ and $\nu$ for different values of loss (channel distance), given the parameters associated to our channel. The values for the detector efficiency, darkcounts, channel loss, and Bob's detection efficiency measured for our underwater polarization channels are given in Table \ref{table3}. These parameters can be used to calculate the values of $\mu$ and $\nu$, which will yield the optimal key rate for different distances. This optimal key rate is plotted in Fig.~\ref{fig:fig4} as the channel length, i.e., attenuation, is increased. The optimal key rate is shown in black accounting for errors resulting from dark counts only. The experimental data points which add the measured QBER to the background errors are shown in red for the channel lengths of 0.5, 10.5, 20.5, and 30.5 meters. The achievable key rates of the experimental points fall below the theoretical line as expected. However, the key rates do closely follow the trend depicted by the theory, suggesting that a channel length near 80 meters could be successfully established with these channel parameters. The key rate analysis for the vector vortex modes follows the exact method demonstrated for the polarization states. The only change is the experimentally observed QBER, which have similar values to those of the polarization channels of the similar length - on the order of 1\% for 10~m - and thus results in similar key rates. For example, the key rate for the 10.5 meter channel of vector vortex modes is 72 Kbps.

\section{Conclusion and Outlook}
We have studied a turbulent underwater channel for quantum communication using polarization and vector vortex modes, beams having spatially structured polarization states. Both the polarization and vector vortex modes maintained their fidelity upon propagation through the channel, resulting in sufficiently low error rates (QBER) to achieve a secure quantum channel. The turbulence - primarily beam wandering - introduces significant challenges with alignment and coupling to single mode fibre. In our channel, the beam path is near the surface of the water for the entire length which results in relatively high turbulence due to the temperature gradient at the air-water surface. Thus channels operating at greater depths may see less turbulence than observed here. Despite these challenges we have shown that both polarization states and spatially structured polarization states can be used in an underwater free-space setting, and for establishing a positive secret key rate for lengths up to 30~meters. The implementation of automated beam tracking equipment would allow one to achieve longer channels approaching the lengths that have been theoretically proposed. In fact, the slower turbulence observed in an underwater channel makes this task of beam correction much easier than in a free-space air environment. We also performed key rate analysis taking into account the parameters measured in our channel. Given these parameters the maximum distance for secure communication would be 80 meters, though this is extremely dependent on the attenuation coefficient of the channel, and Bob's detection efficiency. Improvements in either of these areas would significantly extend the maximum achievable distance.

\section{Acknowledgements} 
We would like to thank Nathalie Brunette, Michel Brassard, Yvan Brunet, and Tony Frade at the Ocean, Coastal and River Engineering Research Centre of National Research Council Canada for all of their efforts in helping us perform this research. This work was supported by Joint Centre for Extreme Photonics (JCEP), Canada Research Chairs (CRC), Canada First Excellence Research Fund (CFREF), and Ontario's Early Researcher Award. A.S. acknowledges the financial support of the Vanier graduate scholarship of the NSERC.

\providecommand{\noopsort}[1]{}

\end{document}